\documentclass[sigplan,nonacm,screen]{acmart}
\usepackage{amsmath}
\usepackage{amsfonts}
\usepackage{amssymb}
\usepackage{amsthm}
\usepackage[normalem]{ulem}
\usepackage{booktabs}
\usepackage{caption}
\usepackage{makecell} %

\usepackage{multirow,bigdelim} %

\usepackage{adjustbox} %

\usepackage{csquotes}

\usepackage{enumitem}
\usepackage{cleveref}

\usepackage{xcolor}
\definecolor{halfgray}{gray}{0.5}
\definecolor{deepblue}{rgb}{0,0,0.5}
\definecolor{deepred}{rgb}{0.6,0,0}
\definecolor{deepgreen}{rgb}{0,0.5,0}

\usepackage{listings}

\lstdefinelanguage{wasm}{
	morekeywords={i32,i64,f32,f64,const,nop,set_local,get_local,tee_local,call,return,call_indirect,get_global,set_global,wrap,shr_s,abs,loop,br,end,block,br_if, drop, select},
	morecomment=[l]{;;},
}
\lstdefinelanguage{JavaScript}{
	keywords={typeof, new, true, false, catch, function, return, null, catch, switch, var, if, in, while, do, else, case, break, let, const, throw},
	comment=[l]{//},
	morecomment=[s]{/*}{*/},
	morestring=[b]',
	morestring=[b]"
}

\makeatletter
\lst@AddToHook{PreSet}{\normallineskiplimit=0pt}
\makeatother

\lstdefinestyle{table}{%
	numbers=none,
	frame=,
	keywordstyle=\bfseries,
	lineskip=1pt,
}

\newcommand{\name}{Wasabi}

\newcommand{\code}[1]{{\ttfamily #1}}

\definecolor{darkred}{rgb}{0.75, 0, 0}

\newcommand{\nbAnalyses}{eight}
\newcommand{\nbBenchmarkPrograms}{32}

\usepackage{floatpag}

\begin{document}

\title{\name{}: A Framework for Dynamically Analyzing WebAssembly}
\date{}

\author{Daniel Lehmann}
\affiliation{\institution{TU Darmstadt, Germany}}
\email{mail@dlehmann.eu}

\author{Michael Pradel}
\affiliation{\institution{TU Darmstadt, Germany}}
\email{michael@binaervarianz.de}

\begin{abstract}
WebAssembly is the new low-level language for the web and has now been implemented in all major browsers since over a year.
To ensure the security, performance, and correctness of future web applications, there is a strong need for dynamic analysis tools for WebAssembly.
Unfortunately, building such tools from scratch requires knowledge of low-level details of the language, and perhaps even its runtime environment.
This paper presents Wasabi, the first general-purpose framework for dynamically analyzing WebAssembly.
Wasabi provides an easy-to-use, high-level API that allows implementing heavyweight dynamic analyses that can monitor all low-level behavior.
The approach is based on binary instrumentation, which inserts calls to analysis functions written in JavaScript into a WebAssembly binary.
Wasabi addresses several unique challenges not present for other binary instrumentation tools, such as
the problem of tracing type-polymorphic instructions with analysis functions that have a fixed type, which we address through an on-demand monomorphization of analysis calls.
To control the overhead imposed by an analysis, Wasabi selectively instruments only those instructions relevant for the analysis.
Our evaluation on compute-intensive benchmarks and real-world applications shows that Wasabi 
(i) faithfully preserves the original program behavior,
(ii) imposes an overhead that is reasonable for heavyweight dynamic analysis (depending on the program and the analyzed instructions, between 1.02x and 163x), and
(iii) makes it straightforward to implement various dynamic analyses, including instruction counting, call graph extraction, memory access tracing, and taint analysis.
\end{abstract}

\maketitle

\section{Introduction}

WebAssembly~\cite{Haas:2017:BWU:3062341.3062363, WasmWebsite} is a new, low-level binary instruction format for the web.
Its core use case is as a compilation target for systems programming languages like C, C++, or Rust. By providing low-level control over the memory layout and by closely mapping to hardware instructions, WebAssembly provides near-native and predictable performance~\cite{herrera2018webassembly, Haas:2017:BWU:3062341.3062363}, unlike managed languages, such as JavaScript and Java.
As of last year, WebAssembly support is enabled in all major browsers~\cite{WasmConsensus}, making it portable across different vendors, architectures, and devices, after prior attempts to establish low-level code on the web have failed, e.g., ActiveX~\cite{ActiveX}, (Portable) Native Client~\cite{NativeClient, Ansel:2011:LSJ:1993498.1993540}, or asm.js~\cite{asmjs}.
To provide some safety guarantees when running untrusted web code, WebAssembly type checks all instructions, strictly separates code from data, verifies that accesses to the linear memory are in-bounds, and offers well-defined interfaces for interaction between modules.
Despite its young age, WebAssembly has already been adopted for various applications, including games\footnote{\url{https://s3.amazonaws.com/mozilla-games/ZenGarden/EpicZenGarden.html}}, cryptography~\cite{Attrapadung:2018:ETH:3196494.3196552}, machine learning~\cite{Gottl:2018:EPT:3208806.3208815}, and medical applications~\cite{doi:10.1177/1176935118771972}.
WebAssembly, it seems, will be a ubiquitous and important instruction set for years to come.

Dynamic analysis tools have a long history of success for languages other than WebAssembly, e.g., to check and understand correctness, security, and performance properties~\cite{Bruening:2011:PMC:2190025.2190067,Seward:2005:UVD:1247360.1247362,UBSan,gcov,JaCoCo,Enck:2014:TIT:2642648.2619091}.
The need for dynamic analysis is particularly strong for highly dynamic languages, such as JavaScript~\cite{Andreasen:2017:SDA:3145473.3106739} and for languages with a lot of low-level control, such as C and C++.
As a compilation target of systems languages and with JavaScript as the host environment, WebAssembly sits exactly at the intersection of these two kinds of languages, making it a prime target for dynamic analysis.

\begin{table*}
	\centering
	\small
	\renewcommand{\arraystretch}{1.1}
	\setlength{\tabcolsep}{7pt}
	\begin{tabular}{l@{\quad}llllll}
		\toprule
		& Pin~\cite{Luk:2005:PBC:1065010.1065034} & Valgrind~\cite{Nethercote:2007:VFH:1250734.1250746} & DiSL~\cite{Marek:2012:DDL:2162049.2162077} & RoadRunner~\cite{Flanagan:2010:RDA:1806672.1806674}
		& Jalangi~\cite{Sen:2013:JSR:2491411.2491447} & \emph{Wasabi}\\\midrule
		(Primary) platform & x86-64 & x86-64 & JVM & JVM & JavaScript & WebAssembly \\
		Instrumentation of \ldots & native binaries & native binaries & byte code & byte code & source code &  binary code\\
		Analysis language & C/C++ & C & Java & Java & JavaScript & JavaScript\\
	API style & \makecell[cl]{instrumentation\\+ callbacks/hooks} & \makecell[cl]{low-level\\instrumentation} & \makecell[cl]{aspect-\\oriented} & event stream & callbacks/hooks & callbacks/hooks\\
		\bottomrule
	\end{tabular}
	\caption{Overview of existing dynamic analysis frameworks and Wasabi.}
	\label{tab:dynamic-analysis-frameworks}
\end{table*}

Creators of a dynamic analysis usually can choose between two options.
One option is to implement the analysis from scratch.
A common strategy is to add instrumentation code to the program, but this requires an in-depth understanding of the instruction set and tools to manipulate it.
Another common strategy is to modify the runtime environment of the program, e.g., a virtual machine.
Unfortunately, such modifications require detailed knowledge of the virtual machine implementation, and they tie the analysis to a particular version of the runtime environment.
Since WebAssembly serves as a compilation target of other languages, source-level instrumentation of these languages might appear to be another possible strategy.
However, typical web applications heavily rely on third-party code, for which source code is unavailable at the client-side.

Instead of implementing a dynamic analysis from scratch, the second option is to build upon general-purpose dynamic analysis frameworks.
\Cref{tab:dynamic-analysis-frameworks} lists some popular frameworks: Pin~\cite{Luk:2005:PBC:1065010.1065034} and Valgrind~\cite{Nethercote:2007:VFH:1250734.1250746} for native programs, DiSL~\cite{Marek:2012:DDL:2162049.2162077} and  RoadRunner~\cite{Flanagan:2010:RDA:1806672.1806674} for JVM byte code, and Jalangi~\cite{Sen:2013:JSR:2491411.2491447} for analyzing JavaScript programs.
Building on an existing framework reduces the overall effort required to build an analysis and enables the analysis author to focus on the design of the analysis itself.
Unfortunately, there currently is no general-purpose dynamic analysis framework for WebAssembly.

This paper presents \emph{Wasabi}, the first general-purpose framework for dynamic analysis of WebAssembly.\footnote{``Wasabi'' stands for \underline{W}eb\-\underline{As}sembly dynamic \underline{a}nalysis using \underline{b}inary \underline{i}nstrumentation.}
Wasabi provides an easy-to-use, high-level API to implement heavyweight analyses that can monitor all low-level behavior.
The framework is based on binary instrumentation, which inserts WebAssembly code that calls into analysis functions in between the program's original instructions.
The analyses themselves are written in JavaScript and implement analysis functions, called \emph{hooks}, to perform arbitrary operations whenever a particular instruction is executed.
To limit the overhead that a dynamic analysis imposes, Wasabi supports \emph{selective instrumentation}, i.e., it instruments only those instructions that are relevant for a particular analysis.

\begin{figure}
	\begin{lstlisting}[language=JavaScript]
const signature = {};
Wasabi.analysis.binary = function(loc, op) {
  switch (op) {
    case "i32.add":
    case "i32.and":
    case "i32.shl":
    case "i32.shr_u":
    case "i32.xor":
      signature[op] = (signature[op] || 0) + 1;
}};
	\end{lstlisting}
	\caption{Cryptominer detection through instruction profiling, as described in a recent WebAssembly analysis~\cite{SEISMIC}.}
	\label{fig:seismic}
\end{figure}

As a simple example of a Wasabi-based analysis, \Cref{fig:seismic} shows our re-implementation of the profiling part of a cryptomining detector~\cite{SEISMIC}.
Unauthorized use of computing resources is detected by monitoring the WebAssembly program and gathering an instruction signature that is unique for typical mining algorithms.
Implementing this analysis in Wasabi takes ten lines of JavaScript, which use the framework's \code{binary} hook to keep track of all executed binary operations.
In contrast, the original implementation is based on a special-purpose instrumentation of WebAssembly that the authors of \cite{SEISMIC} implemented from scratch.
This and more sophisticated analyses (\Cref{sec:eval-analyses}) show that Wasabi allows implementing analyses with little effort.

Apart from being the first dynamic analysis framework for WebAssembly, Wasabi addresses several unique technical challenges.
First, to provide a high-level API for tracking low-level behavior, the approach abstracts away various details of the WebAssembly instruction set.
For example, Wasabi bundles groups of related instructions into a single analysis hook, resolves relative target labels of branch instructions into absolute instruction locations, and resolves indirect call targets to actual functions.
Second, Wasabi transparently handles the interaction of the WebAssembly code to analyze and the JavaScript code that implements the analysis.
A particular challenge is that WebAssembly functions must statically declare fixed parameter types, while some WebAssembly instructions are polymorphic, i.e., they can be executed with different numbers and types of arguments.
To insert hook calls for polymorphic instructions, a different monomorphic variant of the hook must be generated for every concrete combination of argument types.
Wasabi uses \emph{on-demand monomorphization} to automatically create such monomorphic hooks, but only for those type variants that are actually present in the given WebAssembly code.
Third, Wasabi faithfully executes the original program and even preserves its memory behavior, which is useful to implement memory profilers.
To this end, none of the inserted instructions requires access or modification of the program's original memory.
Instead, analyses can track memory operations in JavaScript, i.e., in a separate heap that does not interfere with the WebAssembly heap.

For our evaluation, we implement \nbAnalyses{} analyses on top of Wasabi, including basic block profiling, memory access tracing, call graph analysis, and taint analysis.
We find that writing a new analysis is straightforward and typically takes at most a few dozens of lines of code.
As expected by design, Wasabi faithfully preserves the original program behavior.
The framework instruments large binaries quickly (e.g., 40\,MB in about 15 seconds), and it increases the code size between less than 1\% and 742\%, depending on the program and which instructions shall be analyzed.
The runtime overhead imposed by Wasabi varies between 1.02x and 163x, again depending on the program and instructions-to-analyze, which is in line with existing frameworks for heavyweight dynamic analysis.

\newpage
\medskip
\noindent
In summary, this paper contributes the following:
\begin{itemize}[leftmargin=\parindent]
	\item We present the first general-purpose framework for dynamically analyzing WebAssembly code, an instruction format that is becoming a cornerstone of future web applications.
		
	\item We present techniques to address unique technical challenges not present in existing dynamic analysis frameworks, including on-demand monomorphization of analysis hooks and static resolution of relative branch targets.
	
	\item We show that Wasabi is useful as the basis for a diverse set of analyses, that implementing an analysis takes very little effort, and that the framework imposes an overhead that is reasonable for heavyweight dynamic analysis.
	
	\item We make Wasabi available as open-source, enabling others to build on it: \url{http://wasabi.software-lab.org}

\end{itemize}

\section {Approach}
\label{sec:wasabi}

This section describes our framework for dynamically analyzing WebAssembly programs. We give an overview of Wasabi and the design decisions that have led to it (\Cref{sec:overview}), introduce Mini-Wasm, a WebAssembly core language, (\Cref{sec:wasm-background}), describe the analysis API (\Cref{sec:analysis-api}), and finally present some details of the instrumentation (\Cref{sec:instrumentation}).

\subsection{Overview}
\label{sec:overview}

\begin{figure}
	\includegraphics[width=\linewidth]{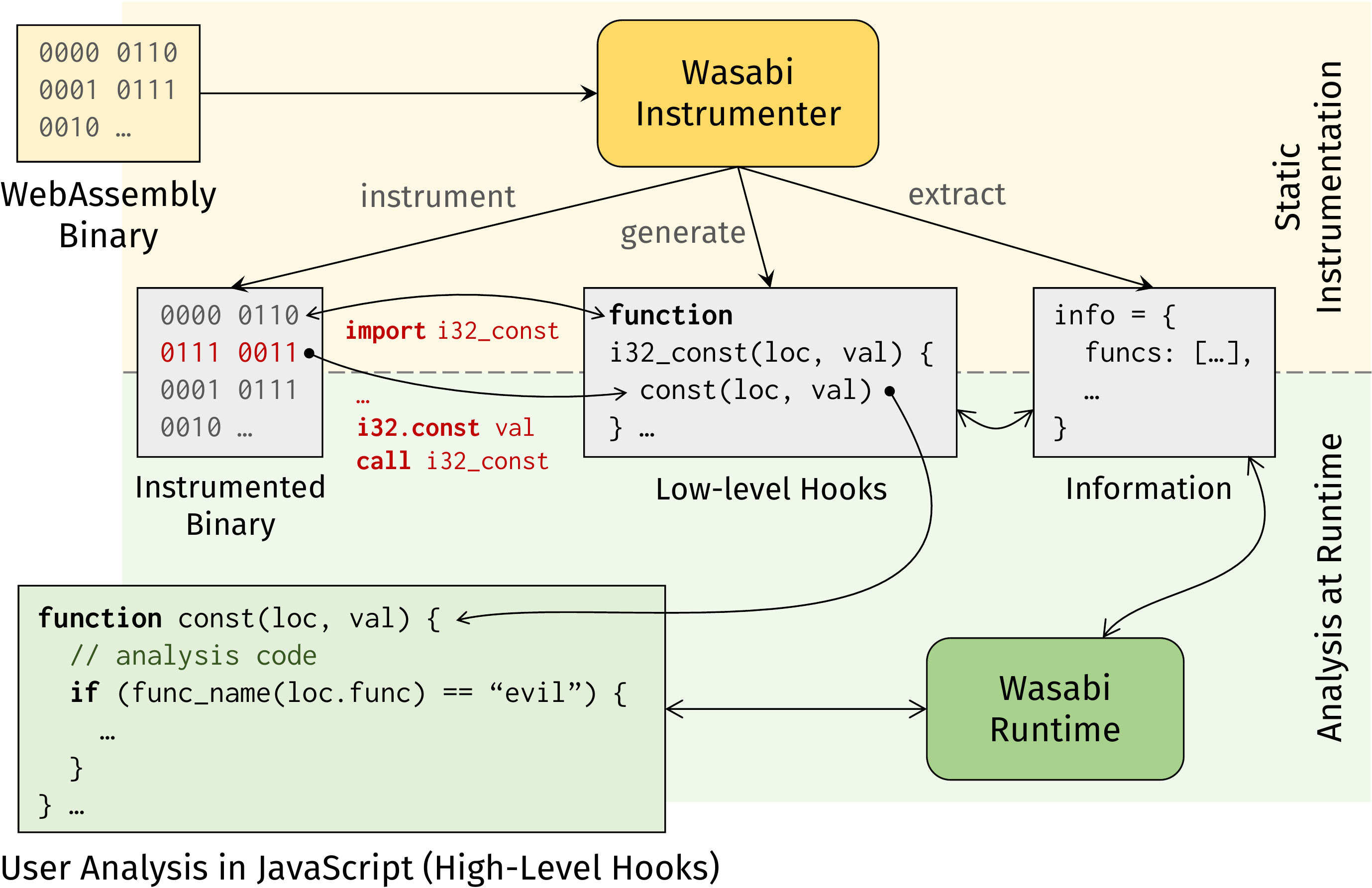}
	\caption{Wasabi's instrumentation and analysis phases.}
	\label{fig:overview}
\end{figure}

\Cref{fig:overview} gives on overview of Wasabi.
The inputs are a WebAssembly binary to analyze (top-left) and a dynamic analysis written in JavaScript (bottom-left).
The rationale for choosing JavaScript as the analysis language is twofold.
First, it is widely used in the web and hence well-known to people interested in dynamic analysis for the web.
Second, JavaScript is the only high-level language that is directly supported by all platforms that currently execute WebAssembly.

The framework has two phases: \emph{static instrumentation} and \emph{analysis at runtime}.
The first phase augments the given WebAssembly binary with instructions that call into the analysis implementation.
To this end, the instructions of the original program are interleaved with calls to \emph{low-level analysis hooks}.
The low-level hooks are implemented in JavaScript code and generated by Wasabi.
At runtime, the low-level hooks then call \emph{high-level analysis hooks}, which the analysis author implements.
The Wasabi runtime provides further information, e.g., the types of functions in the program, to the analysis.

Ahead-of-time binary instrumentation offers three important advantages over alternative designs.
First, it is independent of the execution platform that WebAssembly runs on and robust to changes in current platforms.
Suppose we would instead modify a specific implementation of WebAssembly, e.g., in Firefox, then Wasabi could not analyze programs executed elsewhere.
Moreover, Wasabi would risk to become outdated when the execution platform evolves.
Second, binary instrumentation enables Wasabi to support all languages that compile to WebAssembly, which currently include C/C++~\cite{Zakai:2011:ELC:2048147.2048224}, Rust~\cite{RustWasm}, Go~\cite{GoWasm}, and TypeScript~\cite{Reiser:2017:AJA:3141871.3141873}.
Alternatively, we could rely on source-level instrumentation for these languages, but since the source code of WebAssembly running on websites is often unavailable, this is not an option if we want to support, e.g., security applications like reverse engineering.
Third, ahead-of-time instrumentation avoids runtime overhead compared to instrumenting code during the execution~\cite{Bruening:2003:IAD:776261.776290, Luk:2005:PBC:1065010.1065034, Nethercote:2007:VFH:1250734.1250746}.
Since WebAssembly, in contrast to other binary formats, does not suffer from language features that make ahead-of-time instrumentation inherently difficult, such as self-modifying code or mixed code and data, Wasabi can reliably instrument code ahead-of-time.

\subsection{Mini-Wasm}
\label{sec:wasm-background}

\newcommand{\explanation}[1]{\tag*{\raisebox{.5pt}{\normalsize\color{gray}{#1}}}}
\newcommand{\nonterm}[1]{\text{\rmfamily\itshape #1}}
\newcommand{\literal}[1]{\text{\ttfamily\bfseries\footnotesize #1}}
\newcommand{\idx}[1]{\nonterm{idx\textsubscript{#1}}}
\newcommand{\type}[1]{\nonterm{type\textsubscript{#1}}}
\newcommand{\s}{^{\kern1pt*}}
\newcommand{\?}{^{\kern1pt\raisebox{-1pt}{\scriptsize?}}}
\newcommand{\+}{^{\kern1pt+}}

\begin{figure}
	\vspace{-12pt}
	\small
	\setlength{\jot}{.2pt}
	\begin{align*}
		\explanation{Module and sections}\\
		\nonterm{module} ::=&\ \nonterm{function}\s\ \nonterm{global}\s\ \nonterm{start}\?\ \nonterm{table}\?\
		\nonterm{memory}\?\\[\medskipamount]
		\nonterm{function} ::=&\ \type{func}\enspace (\nonterm{import} \mid\nonterm{code})\enspace  \nonterm{export}\s\\
		\nonterm{global} ::=&\ \type{val}\enspace (\nonterm{import}\ \mid \nonterm{init})\enspace \nonterm{export}\s\\
		\nonterm{start} ::=&\ \idx{func}\\
		\nonterm{table} ::=&\ \nonterm{import}\?\ \idx{func}\kern-1pt\s\ \nonterm{export}\s\\
		\nonterm{memory} ::=&\ \nonterm{import}\?\ \literal{\mdseries byte}\kern-1pt\s\ \nonterm{export}\s\\[\medskipamount]
		\nonterm{import} ::=&\ \literal{\mdseries "name"},\quad\nonterm{export} ::= \literal{\mdseries "name"}\\
		\nonterm{code} ::=&\ (\literal{local}\ \type{val})\kern-1pt\s\ \nonterm{instr}\s\\
		\nonterm{init} ::=&\ \nonterm{instr}\s\\[\bigskipamount]
		\explanation{Instructions}
		\nonterm{instr} ::=&\ \literal{nop} \mid \literal{drop} \mid \literal{select}
		\\
		\mid&\ \type{val}\literal{.const}\ \nonterm{value} \mid \nonterm{unary} \mid \nonterm{binary}
		\\
		\mid&\ \type{val}\literal{.load} \mid \type{val}\literal{.store} \mid \literal{memory.grow}
		\\
		\mid&\ (\literal{set}\,|\,\literal{get}\,|\,\literal{tee})\literal{\_local}\ \idx{local} 
		\mid (\literal{set}\,|\,\literal{get})\literal{\_global}\ \idx{global}
		\\
		\mid&\ \literal{call}\ \idx{func} \mid \literal{call\_indirect}\ \type{func} \mid \literal{return}
		\\
		\mid&\ \literal{block} \mid \literal{loop} \mid \literal{if} \mid \literal{else} \mid \literal{end}
		\\
		\mid&\ \literal{br}\ \nonterm{label} \mid \literal{br\_if}\ \nonterm{label} \mid \literal{br\_table}\ \nonterm{label}\+
		\\
		\nonterm{unary} ::=&\ \literal{i32.eqz} \mid ... \mid \literal{f32.neg} \mid ... \mid \literal{f32.convert\_s/i32} \mid ...
		\\
		\nonterm{binary} ::=&\ \literal{i32.add} \mid ... \mid \literal{i32.eq} \mid ...
		\\[\bigskipamount]
		\explanation{Types, labels, indices}
		\type{val} ::=&\ \literal{i32} \mid \literal{i64} \mid \literal{f32} \mid \literal{f64}\\
		\type{func} ::=&\ \type{val}\s\kern-1pt \rightarrow \type{val}\s\\[\smallskipamount]
		\nonterm{label} \in\ &\mathbb{N},\quad\idx{func\,$\mid$\,global\,$\mid$\,local} \in \mathbb{N}
	\end{align*}
	\vspace{-15pt}
	\caption{Abstract syntax of Mini-Wasm.}
	\label{fig:mini-wasm}
	\vspace{-10pt}
\end{figure}

For a self-contained and concise presentation, we now introduce \emph{Mini-Wasm}, a simplified core of WebAssembly.
The Wasabi implementation supports the entire WebAssembly language.
\Cref{fig:mini-wasm} shows the grammar of Mini-Wasm modules.
A \emph{module} corresponds to a single binary file and contains \emph{functions}, \emph{global} variables, an optional \emph{start} function, and at most one table and memory.
Each of these have a name only when they are imported or exported, and otherwise are referenced by integer indices.
The \emph{table} maps indices to functions and is used for indirect calls, e.g., to implement function pointers or virtual calls.
Similar to native programs, and unlike in managed languages with garbage collection, WebAssembly \emph{memory} is a linear sequence of bytes, which can be increased at runtime with \code{memory.grow}. 

WebAssembly execution is a stack machine with per-function locals, similar to the JVM~\cite{JvmSpec}.
One distinctive feature of WebAssembly, which is relevant for Wasabi, is how control-flow is encoded.
Unlike in the JVM or native code, instructions are structured into well-nested, implicitly labeled blocks.
Instead of unrestricted gotos that directly jump to an instruction offset, branch instructions can only target blocks in which they are enclosed.
The destination block is referenced by a non-negative integer label, where zero indicates the immediately enclosing block.
Depending on the block type, the next executed instruction is either the first one inside the block (for \code{loop} blocks, rendering the branch a backward jump) or the next instruction after the block's matching \code{end} instruction (for \code{block}, \code{if}, and \code{else} blocks, thus a forward jump).
For example, in \Cref{fig:wasm-cf}, the label \code{1} at line~\ref{line:branch} references the block at line~\ref{line:block-begin}, and hence is a jump forward to line~\ref{line:next-instruction}.
\Cref{sec:instrumentation} describes how Wasabi handles this control-flow encoding.

WebAssembly is statically type checked and knows four primitive types for globals, locals, and stack values: 32 and 64 bit integers (\code{i32}, \code{i64}), and single and double precision floats (\code{f32}, \code{f64}).
Many WebAssembly instructions are polymorphic in the sense that the input and result types vary depending on the context in which the instruction is executed.
For example, \code{call} and \code{return} pop and push different types depending on the function type of the called and current function, respectively.
Similarly, the instruction types for accesses to locals and globals depend on the referenced local and global variable.

\begin{figure}
	\begin{lstlisting}
block       <---------, $\label{line:block-begin}$
    block             |  
        get_local 0   |
        br_if 1    ---' ;; block reference by label$\label{line:branch}$
        ;; next instruction if local #0 == false
    end
end ;; matching end for first block
;; next instruction if local #0 == true $\label{line:next-instruction}$
	\end{lstlisting}
	\caption{Structured control-flow in WebAssembly.}
	\label{fig:wasm-cf}
\end{figure}

For \code{drop}, which removes the current stack top, and \code{select}, which pushes one of two values depending on a condition, the types cannot be simply looked up in the module, but depend on the previously executed code.
For example, a \code{drop} following an \code{i32.const} has \code{i32} as input type, whereas  a \code{drop} following a call that returns an \code{f64} value has \code{f64} as input type.
The many possible typed instructions pose a challenge for generating Wasabi's hooks, which we explain along with our solution in \Cref{sec:instrumentation}.

\subsection{Analysis API}
\label{sec:analysis-api}

Wasabi offers analysis authors an API with hooks to be implemented by the analysis.
The API is both powerful enough to enable arbitrary dynamic analyses and high-level enough to spare the analysis author dealing with irrelevant details.
\Cref{tab:hooks} shows the hooks,\footnote{For brevity, the table leaves out five additional hooks that Wasabi supports, \code{start}, \code{nop}, \code{unreachable}, \code{if}, and \code{memory\_size}, for a total of 23 hooks.} along with their arguments and types.
The hooks can be roughly clustered into six groups:
Instructions related to stack manipulation (\code{const}, \code{drop}, \code{select}),
operations (\code{unary}, \code{binary}),
accesses and management of register and memory (\code{local}, \code{global}, \code{load}, \code{store},  \code{memory\_grow}),
function calls (\code{call\_pre}, \code{call\_post}, \code{return}),
control flow (\code{br}*), and
blocks (\code{begin}, \code{end}).
Each hook implementation receives details about the respective instruction, e.g., its inputs and outputs, as well as the code location of the instruction.

\begin{figure}
	\centering
	\includegraphics[width=.38\linewidth]{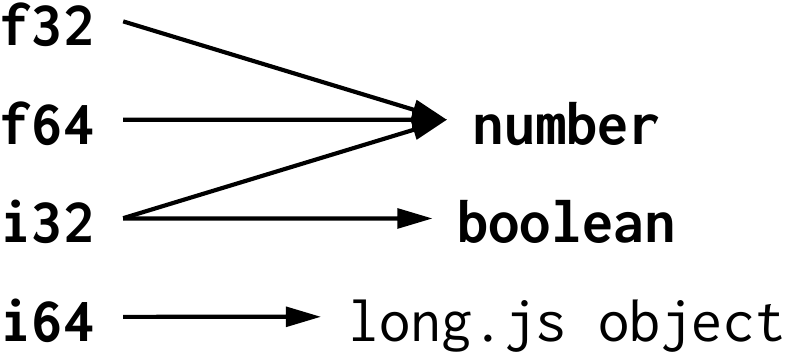}
	\caption{Mapping of WebAssembly types to JavaScript.}
	\label{fig:type-mapping}
\end{figure}

The API is designed to ensure four important properties.
\begin{itemize}[leftmargin=\parindent]
	\item \emph{Full instruction coverage}. It covers all WebAssembly instructions and provides all their inputs and results to the analysis.
	This property is crucial to implement arbitrary dynamic analyses that can observe all runtime behavior.
	We describe in \Cref{sec:instrumentation} how selective instrumentation limits the costs to be paid for this flexibility.

	\item \emph{Grouping of instructions}. The API groups related WebAssembly instructions into a single hook, which significantly reduces the number of hooks analyses must implement.
	Providing one hook per instruction to the analysis would require a huge number of hooks (e.g., there are 123 numeric instructions alone), whereas Wasabi's API provides 23 hooks only.
	To distinguish between instructions, if necessary, the hooks receive detailed information as arguments.
	For example, the \code{binary} hook receives an \code{op} argument that specifies which binary operation is executed.
	To hide the various variants of polymorphic instructions from analyses authors, Wasabi also maps all variants of the same kind of instruction into a single hook.
	For example, the \code{call} instruction can take different numbers and types of arguments, depending on the called function, which are represented in the hook as an array of varying length.
	
	\item \emph{Pre-computed information}.
	Wasabi provides pre-computed information along with some hooks because the runtime values on their own are not informative enough for an analysis.
	For example, the three branch-related hooks receive \code{target} objects that contain the statically resolved, absolute location of the next instruction that will be executed, if the branch is taken, alongside the low-level relative branch label.
	Similarly for indirect calls, Wasabi resolves the runtime table index to the actually called function.
	
	\item \emph{Faithful type mappings}. Finally, the API faithfully maps typed values from WebAssembly to JavaScript.
	\Cref{fig:type-mapping} shows the four primitive WebAssembly types and how they are represented without loss of precision in a Wasabi analysis.
	\code{i32}, \code{f32}, and \code{f64} are represented as a JavaScript \code{number}.
	Since JavaScript has no native support for 64-bit integers, Wasabi transparently maps them to \code{long.js}\footnote{\url{https://github.com/dcodeIO/long.js}} objects.
	Conditions, which are \code{i32}s with value 0 or 1 in WebAssembly, are mapped to JavaScript's  \code{boolean}s.
\end{itemize}

\begin{table}
	\lstset{style=table,basicstyle=\small\ttfamily,morekeywords={unary,binary,load,store,memory_grow,local, global,call_pre,call_post,begin,end,br_table}}
	\small
	\begin{adjustbox}{center}
		\begin{tabular}{@{\hskip2pt}l@{\hskip-35pt}r@{\hskip2pt}}
			\toprule
			\normalsize Hook Name & \normalsize Arguments and Types\\
			\midrule
	
			\lstinline|const|\kern1pt/\kern.5pt\lstinline|drop| & \lstinline|value|: $\type{val}$\\
			\lstinline|select| & \lstinline|condition|: \lstinline|boolean|, \lstinline|first|: $\type{val}$, \lstinline|second|: $\type{val}$\medskip\\
			
			\lstinline|unary| &  \lstinline|op|, \lstinline|input|: $\type{val}$, \lstinline|result|: $\type{val}$\\
			\lstinline|binary| & \lstinline|op|, \lstinline|first|: $\type{val}$, \lstinline|second|: $\type{val}$, \lstinline|result|: $\type{val}$\\
			\lstinline|local|\,/\kern.5pt\lstinline|global| & \lstinline|op|, \lstinline|index|: \lstinline|number|, \lstinline|value|: $\type{val}$\\
			\enspace\emph{where} & \lstinline|op|: instruction \lstinline|string|, e.g., \code{i32.add} or \code{get\_local}\medskip\\
			
			\lstinline|memory_grow| & \lstinline|delta|: \lstinline|number|, \lstinline|previousSize|: \lstinline|number|\\
			\lstinline|load|\,/\kern.5pt\lstinline|store| & \lstinline|op|, \lstinline|memarg|, \lstinline|value|: $\type{val}$\\
			\enspace\emph{where} & \lstinline|memarg|: \lstinline|{addr|: \lstinline|number|, \lstinline|offset|: \lstinline|number}|\medskip\\
			
			\lstinline|call_pre| & \lstinline|func|: \lstinline|number|, \lstinline|args|, \lstinline|tableIndex|: (\lstinline|number| | \lstinline|null|)\\
			\enspace\emph{where} & \lstinline|args|: \code{[}$\type{val}$\code{]}, \lstinline|tableIndex| \lstinline|==| \lstinline|null| iff direct call\\
			\lstinline|call_post|\kern1pt/\kern.5pt\lstinline|return| & \lstinline|results|: \code{[}$\type{val}$\code{]}\medskip\\
			
			\lstinline|br| & \lstinline|target|\\
			\lstinline|br_if| & \lstinline|target|, \lstinline|condition|: \lstinline|boolean|\\
			\lstinline|br_table| & \lstinline|table|: \lstinline|[target]|, \lstinline|tableIndex|: \lstinline|number|\\
			\enspace\emph{where} & \lstinline|target|: \lstinline|{label|: \lstinline|number|, \lstinline|location|: \lstinline|location}|\medskip\\
			
			\lstinline|begin| & \lstinline|type|\\
			\lstinline|end| & \lstinline|type|, \code{begin}: \lstinline|location|\\
			\enspace\emph{where} & \lstinline|type|: \lstinline|string| $\in \lbrace\texttt{function}, \texttt{block}, \texttt{loop}, \texttt{if}, \texttt{else}\rbrace$\medskip\\
			
			\emph{every hook} & \lstinline|location|: \lstinline|{func|: \lstinline|number|, \lstinline|instr|: \lstinline|number}|, \ldots\\
			
			\bottomrule
		\end{tabular}
	\end{adjustbox}
	\caption{API of the high-level analysis hooks.}
	\label{tab:hooks}
\end{table}

\medskip
\noindent
The API gives analysis authors the power to implement sophisticated dynamic analyses with little effort.
In particular, it is straightforward to implement memory shadowing~\cite{Nethercote:2007:VFH:1250734.1250746,DBLP:conf/cgo/ZhaoBA10}, a feature useful, e.g,. for tracking the origin of undesired values or for taint analysis.
To associate some meta-information with a memory value, all an analysis must do is to maintain a map of memory locations to meta-information, and to update the meta-information on memory-related instructions.
One of our example analyses (\Cref{sec:eval-analyses}) is a taint analysis that implements memory shadowing in this way.

\subsection{Binary Instrumentation}
\label{sec:instrumentation}

The following presents the core of Wasabi: its binary instrumentation component, which inserts code that eventually calls the high-level analysis hooks described in the previous subsection.
We first describe the instrumentation of individual WebAssembly instructions (\Cref{sec:instrInstr}) and how Wasabi reduces overhead via selective instrumentation (\Cref{sec:selectiveInstr}).
Then, we highlight four instrumentation challenges that are unique to WebAssembly and describe how Wasabi addresses them (Sections~\ref{sec:mono} to~\ref{sec:i64}).

\newenvironment{explaintable}{\small\setlength\extrarowheight{-.5pt}\begin{tabular}{@{}l}}{\end{tabular}}
\newcommand{\tableheading}[1]{& \multicolumn{3}{@{}l@{}}{\small #1:}}

\begin{table*}
	\thisfloatpagestyle{empty}
	\begin{adjustbox}{center}
	\lstset{style=table}
	\begin{tabular}{@{}rl@{}l@{\enspace}l@{\hspace*{-12pt}}}
\toprule
& \multicolumn{2}{l}{\hspace{2.7em} Instructions} & \multirow{2}{*}{\makecell{Explanations and other changes made to the module}}\\
& Original\hspace{1.5em}$\Rightarrow$\hspace*{1.5em}& Instrumented &\\
\specialrule{\lightrulewidth}{.4ex}{6pt}
\tableheading{Constants ($\type{val}$\code{\bfseries.const}), similarly simple instructions that only produce a value}\\[-1pt]
1 &
\begin{lstlisting}
i32.const $\nonterm{value}$


\end{lstlisting}
&
\begin{lstlisting}
i32.const $\nonterm{value}$
i32.const $\nonterm{value}$
call $\idx{\kern1pt\code{hooks.i32.const}}$
\end{lstlisting}
&
\begin{explaintable}
original instruction\\
duplicate constant value for the hook\\
instrumentation hook, also needs to be added as function import to module\\
\end{explaintable}
\\
\midrule
\tableheading{General instructions (unary, binary, load, store)}\\[-1pt]
2 &
\begin{lstlisting}

f32.abs




\end{lstlisting}
&
\begin{lstlisting}
tee_local $\idx{temp input local}$
f32.abs
tee_local $\idx{temp result local}$
get_local $\idx{temp input local}$
get_local $\idx{temp result local}$
call $\idx{\kern1pt\code{hooks.f32.abs}}$
\end{lstlisting}
&
\begin{explaintable}
	store instruction input(s) into freshly generated local(s)\\
	\\
	store instruction result into freshly generated local\\
	\rdelim\}{2}{*}[\,push inputs and results on stack as hook arguments]\\
	\\
	one low-level hook per instruction\\
\end{explaintable}
\\
\midrule
\tableheading{Calls (\code{call} and \code{call\_indirect}), similarly also \code{return}s}\\[-1pt]
3 &
\begin{lstlisting}




call $\idx{func}$



\end{lstlisting}
&
\begin{lstlisting}
tee_local $\idx{temp input local}$
i32.const $\idx{func}$
get_local $\idx{temp input local}$
call $\idx{\kern1pt\code{hooks.call\_pre\_}(\type{local})*}$
call $\idx{func}$
tee_local $\idx{temp result local}$
get_local $\idx{temp input local}$
call $\idx{\kern1pt\code{hooks.call\_post\_}(\type{local})*}$
\end{lstlisting}
&
\begin{explaintable}
	store call argument(s) in freshly generated local(s)\\
	pass index of called function to hook\\[1pt]
	\rdelim\}{2}{*}[\,pass stored inputs to monomorphized \code{call\_pre} hook]\\
	\\[1pt]
	\\
	store call result(s) in freshly generated local(s)\\
	\rdelim\}{2}{*}[\,pass stored results to monomorphized \code{call\_post} hook]\\
	\\
\end{explaintable}
\\
\midrule
\tableheading{Polymorphic \code{drop} and \code{select} instructions}\\[-1pt]
4 &
\begin{lstlisting}
$\ldots$
$\textrm{(preceding code)}$
$\ldots$
drop
\end{lstlisting}
&
\begin{lstlisting}
$\ldots$
$\textrm{(preceding code)}$
$\ldots$
call $\idx{\kern1pt\code{hooks.drop\_}\type{val}}$
\end{lstlisting}
&
\begin{explaintable}
	type check all instructions to keep track of abstract stack\\
	here: assuming preceding instructions produce a stack top value of type \type{val},\\
	then the following \code{drop} has type \code{[\type{val}]} $\rightarrow$ \code{[]}\\
	matching monomorphic hook call is inserted (consumes stack top in place of \code{drop})
\end{explaintable}
\\[5pt]
\midrule
\tableheading{Blocks\kern.5pt/structured control-flow (\code{block}, \code{loop}, \code{if}, \code{else}, \code{end}) and branches (\code{br}, \code{br\_if}, \code{br\_table})}\\[-1pt]
5 &
\begin{lstlisting}
$\nonterm{label}\!:$ block

  $\nonterm{otherlabel}\!:$ loop

    $...$





    br 1 $(\kern-.5pt\nonterm{label}\kern.5pt)$
    $...$

  end

end
\end{lstlisting}
&
\begin{lstlisting}
$\nonterm{label}\!:$ block
  call $\idx{\code{hooks.begin\_block}}$
  $\nonterm{otherlabel}\!:$ loop
    call $\idx{\code{hooks.begin\_loop}}$
    $...$
    i32.const 1 $(\kern-.5pt\nonterm{label}\kern.5pt)$
    i32.const $\text{\rmfamily resolve}(\kern-.5pt\nonterm{label}\kern.5pt)$
    call $\idx{\code{hooks.br}}$
    call $\idx{\code{hooks.end\_loop}}$
    call $\idx{\code{hooks.end\_block}}$
    br 1 $(\kern-.5pt\nonterm{label}\kern.5pt)$
    $...$
    call $\idx{\code{hooks.end\_loop}}$
  end
  call $\idx{\code{hooks.end\_block}}$
end
\end{lstlisting}
&
\begin{explaintable}
\\[-5pt] %
\nonterm{label} is implicit and not encoded in the Wasm binary\\
every block type (\code{block}/\code{loop}/\code{if}/\code{else}) has own low-level \code{begin} hook\\
\\
loop \code{begin} hook is called once per iteration\\
\\
\enquote{raw} (i.e., unresolved relative) target label is passed to hook as integer\\
resolved (at instrumentation time) label to next executed instruction is also passed\\
branch hooks must come before the instruction\\
\rdelim\}{2}{*}[\,explicitly call \code{end} hooks of all \enquote{traversed} blocks, for dynamic block nesting]\\
\\[2pt]
\\
(not shown:) end hooks receive location of the end and of the matching block begin\\
every block type (\code{block}/\code{loop}/\code{if}/\code{else}) has own \code{end} hook (cf. \code{begin\_*} hook)\\
\\
\\
\\
\end{explaintable}
\\
\midrule
\tableheading{Instructions with \code{i64} inputs or results, value is split for passing to hook}\\[-1pt]
6 &
\begin{lstlisting}
i64.const $\nonterm{value}$







\end{lstlisting}
&
\begin{lstlisting}
i64.const $\nonterm{value}$
i64.const $\nonterm{value}$
i32.wrap/i64
i64.const $\nonterm{value}$
i64.const 32
i64.shr_s
i32.wrap/i64
call $\idx{\kern1pt\code{hooks.i64.const}}$
\end{lstlisting}
&
\begin{explaintable}
	if instruction has side-effects, its result is duplicated via a local instead (but const here)\\[-1pt]
	\rdelim\}{2}{*}[\,push lower 32-bit half of \code{i64} value as \code{i32} on stack]\\
	\\[2pt]
	\rdelim\}{4}{*}[\,shift upper 32-bit half of \code{i64} value to right, then push as \code{i32} on stack]\\
	\\
	\\
	\\
	cannot pass \code{i64} values to hooks, so they take a tuple of (\code{i32}, \code{i32}) instead\\
\end{explaintable}
\\
\bottomrule
	\end{tabular}
	\end{adjustbox}
	\caption{Instrumentation of (a subset of all) WebAssembly instructions. Every hook also takes two \code{i32}s that represent the original instruction's location. For brevity, the corresponding \code{i32.const} instructions are omitted in the table.}
	\label{tab:instrumentation}
\end{table*}

\subsubsection{Instrumentation of Instructions}
\label{sec:instrInstr}

To keep track of all instructions that occur during the execution, Wasabi inserts for each instruction a call to an analysis hook.
\Cref{tab:instrumentation} illustrates the instrumentation for a subset of all instructions.
Row~1 shows the simplest case: a \code{const} instruction that pushes an immediate value on the stack.
The instrumentation adds a call to the corresponding hook.
Since the hook receives the value produced by the \code{const} instruction as an argument, the value is pushed once more on the stack prior to the call.
After the call to the hook, the stack will be the same as in the original, uninstrumented program.

Row~2 of \Cref{tab:instrumentation} shows an instruction that takes inputs and produces results.
To pass both to the inserted hook call, we need to duplicate values on the stack.
For this purpose, Wasabi generates a fresh local of the appropriate type and writes the current stack top to this local with \code{tee\_local}.
Before the hook call, the inserted code retrieves the stored input and its result with \code{get\_local}.
Row~3 illustrates how Wasabi instruments \code{call} instructions.
In contrast to other instructions, we surround the original instruction with two calls into the analysis, so that an analysis author can execute analysis behavior before and after the call.

All inserted calls go to JavaScript functions that are imported into the WebAssembly binary.
These imported functions are not yet the high-level hooks from \Cref{sec:analysis-api}, but low-level hooks that are automatically generated by Wasabi.
There are several reasons for this indirection.
First, it allows Wasabi to map typed WebAssembly instructions to untyped JavaScript hooks in a seamless way (\Cref{sec:mono}).
Second, it helps providing information that is useful in high-level hooks but not available at the current instruction (\Cref{sec:resolving}).
Third, \Cref{sec:blocks} shows that Wasabi sometimes also calls other hooks at runtime, because the necessary information which hooks to call is available only then.
Finally, the low-level hooks can convert values before passing them to the high-level hooks (\Cref{sec:i64}).
All of these issues can be solved by automatically generated low-level hooks that are hidden from analysis authors.

\subsubsection{Selective Instrumentation}
\label{sec:selectiveInstr}

Not every analysis uses all of the hooks provided by the API from \Cref{sec:analysis-api}.
To reduce both the code size and the runtime overhead of the instrumented binary, Wasabi supports \emph{selective instrumentation}.
That is, only those kinds of instructions are instrumented that have a matching high-level hook in the given analysis.
Wasabi ensures that the instrumentation for different kinds of instructions are independent of each other, so that instrumenting only some instructions still correctly reflects their behavior.
Sections~\ref{sec:eval-code-size} and \ref{sec:eval-runtime-overhead} show that selective instrumentation significantly reduces code size and runtime overhead.

\subsubsection{On-demand Monomorphization}
\label{sec:mono}

An interesting challenge for the instrumentation comes from static typing in WebAssembly.
While there are polymorphic instructions, WebAssembly functions, including our hooks, must always be declared with a fixed, monomorphic type.
For polymorphic instructions, Wasabi cannot simply generate one hook per kind of instruction:
Consider \code{drop} with the polymorphic instruction type [{\small$\alpha$}]\,$\rightarrow$\,[\,].
Inserting a call to the same hook after each \code{drop} is not possible, because the hook's function type would then be polymorphic.
Instead, Wasabi generates multiple monomorphic variants of a polymorphic hook and inserts a call to the appropriate monomorphic low-level hook.\footnote{This strategy is similar to the compilation of generic functions in Rust or instantiation of function templates in C++~\cite{klabnik2018rust, Vandevoorde:2002:CT:579240}.}

For many polymorphic instructions, determining which monomorphic hook variant to call is straightforward. For example, the instruction type of \code{set\_global} depends only on the type of the referenced variable.
The types of \code{drop} and \code{select}, however, cannot be simply looked up.
Instead, as shown in row~4 of \Cref{tab:instrumentation}, their type depends on all preceding instructions. Wasabi thus performs full type checking during instrumentation, that is, it keeps track of the types of all values on the stack~\cite{Haas:2017:BWU:3062341.3062363, Watt:2018:MVW:3176245.3167082}. When the \code{drop} in the last line of the example is encountered, its input type is equal to the top of the abstract stack and Wasabi can insert the call to the matching monomorphic low-level hook.

While creating monomorphic variants of hooks yields type-correct WebAssembly code, doing so eagerly leads to an explosion of the required number of monomorphic hooks.
Since functions can have an arbitrary number of arguments and results\footnote{Strictly speaking, functions in the binary format 1.0 have at most one result, but the formal semantics already support multiple return values~\cite{Haas:2017:BWU:3062341.3062363}.}, the number of monomorphic hooks for \code{call}s and \code{return}s is even unbounded.
One way to address this problem would be to set a heuristic limit, e.g., by generating hooks for calls with up to ten arguments.
However, the resulting $4^{10} = 1,048,576$ call-related hooks would cause unnecessary binary bloat and may still fail to support all calls.

Instead, Wasabi generates monomorphic hooks on-demand only for instructions and type combinations that are actually present in the given binary.
We call this approach \emph{on-demand monomorphization} of hooks.
During instrumentation, Wasabi maintains a map of already generated low-level hooks.
If a required hook, e.g., for a \code{call} instruction with type [\code{i32}]\,$\rightarrow$\,[\code{f32}], is present in the map, the function index of the hook is returned.
Otherwise, Wasabi generates the hook and updates the map.
Our evaluation shows that on-demand monomorphization significantly reduces the number of low-level hooks, and hence the code size, compared to the eager approach described above.

\subsubsection{Resolving Branch Labels}
\label{sec:resolving}

As described in \Cref{sec:wasm-background}, WebAssembly relies on structured control-flow, a feature not present in other low-level instruction sets.
An interesting challenge that arises from structured control-flow is how and when to resolve the destination of branches.
Row~5 of \Cref{tab:instrumentation} illustrates the problem with a few control-flow-related instructions.
The \code{br} instruction jumps to a destination referenced by a relative integer label \code{1}.
However, passing this label to the high-level dynamic analysis API would be of limited use, because 
without additional static information (namely the surrounding blocks), the dynamic analysis cannot resolve the label to a code location.

\begin{figure}
	\centering
	\includegraphics[width=.45\linewidth]{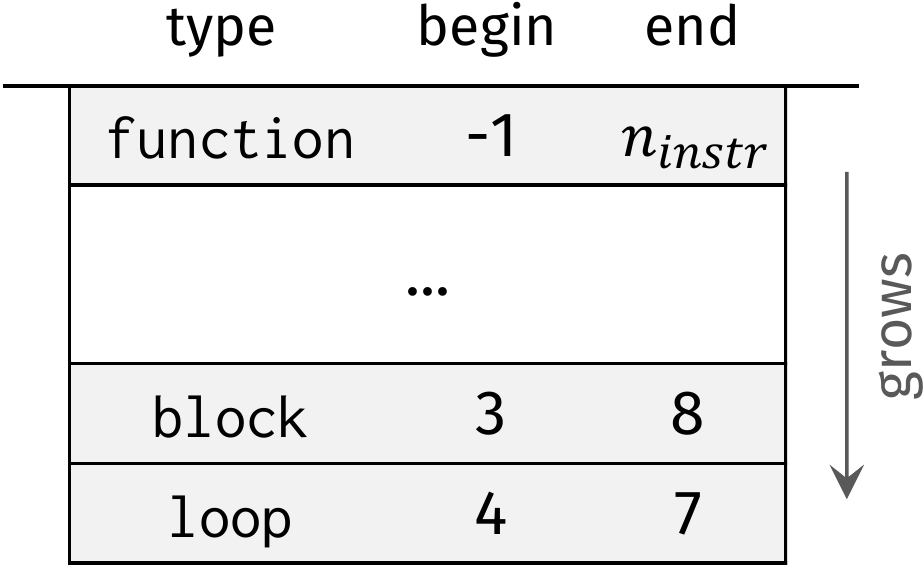}
	\caption{Abstract control stack at the \code{br} branch instruction in row~5 of \Cref{tab:instrumentation} (assuming the example is preceded by four other instructions).}
	\label{fig:control-stack}
\end{figure}

To enable analysis authors to reason about branch destinations without implementing their own static analysis, Wasabi resolves branch labels during the instrumentation and passes the resulting absolute instruction locations to the high-level API.
To resolve branch labels, Wasabi keeps track of an \emph{abstract control stack} while instrumenting WebAssembly code.
Whenever the instrumentation enters a new block, an element is pushed to the control stack, consisting of the block type (\code{function}, \code{block}, \code{loop}, \code{if}, or \code{else}), the location of the block begin, and the location of the matching \code{end} instruction.
Whenever the instrumentation encounters the end of a block, the topmost entry is popped of the control stack.
As an example, \Cref{fig:control-stack} shows the control stack for the code in row~5 of \Cref{tab:instrumentation}. 

Given the abstract control stack, Wasabi can determine during instrumentation what code location a branch, if taken, will lead to.
At every branch to a label $n$, Wasabi queries the control stack for its $n+1$-th entry from the top, to determine the targeted block, and then computes the location of the next instruction from the block type and the locations of the \code{begin} and \code{end} instructions.
This absolute instruction location is then given as an argument to the branch hook, as shown in the example in \Cref{tab:instrumentation}.

\subsubsection{Dynamic Block Nesting}
\label{sec:blocks}

Another control-flow-related challenge is about observing the end of the execution of a block.
Some analyses may want to observe the block nesting at runtime, i.e., to perform some action when a block is entered and left.
For this purpose, Wasabi offers the high-level \code{begin} and \code{end} hooks (\Cref{sec:analysis-api}).
The example in row~5 of \Cref{tab:instrumentation} shows that our instrumentation adds the respective hook calls (e.g., \code{call} \idx{\code{hooks.begin\_block}} and \code{call} \idx{\code{hooks.end\_block}}) at the beginning of a block and before the matching \code{end}.

Unfortunately, \code{br}anching or \code{return}ing will jump out of a block and over the inserted \code{end} hook calls. Consider the last two calls to \code{hooks.end\_loop} and \code{hooks.end\_block} in \Cref{tab:instrumentation}. They are not executed because the earlier \code{br} \code{1} directly transfers control to after the enclosing block. To account for that, Wasabi adds additional calls before each \code{br}anch and \code{return} that invoke every end hook of the blocks that will be \enquote{traversed} during the jump. That is, as the example shows, Wasabi inserts calls to the \code{end} hooks for the two enclosing blocks prior to the \code{br} \code{1} instruction. Again, the \emph{control stack} can tell us which end hooks need to be called, namely all between the current block (stack top, inclusive) and the branch target block (also inclusive). For example, in \Cref{fig:control-stack}, the instrumented code calls the \code{loop} and \code{block} end hooks. For a \code{return} it would be all blocks on the block stack up to and including the \code{function} block.

For conditional branches (\code{br\_if}), we call the \code{end} hooks for traversed blocks only if the branch is actually taken.
Similarly, for multi-way branches (\code{br\_table}), which branch is taken (and thus which blocks are left) is known only at runtime.
Thus, the instrumentation statically extracts the list of ended blocks for every branch table entry and stores this information.
Inside the low-level hook for \code{br\_table}, one of the stored branch table entries will then be selected, before calling the corresponding \code{end} hooks at runtime.

\subsubsection{Handling \code{i64} Values}
\label{sec:i64}

As mentioned in \Cref{sec:wasm-background}, \code{i64} values cannot be passed to JavaScript functions (and thus our hooks), since JavaScript has only double precision float numbers.
To nevertheless enable dynamic analyses to faithfully observe all runtime values, including \code{i64} values, Wasabi splits a 64-bit integer into two 32-bit integers to pass them to JavaScript.
For every \code{i64} stack value (either produced by a \code{const} or by any other instruction), we thus insert instrumentation as shown in row~6 of \Cref{tab:instrumentation}.
The inserted code duplicates the \code{i64} value twice, from the first of which only the lower 32 bits are extracted and the second of which is shifted to result in the upper 32 bits.
Both \code{i32} values can then be passed to the hook in question.
On the JavaScript side, the low-level hook joins the two 32-bit values into a \code{long.js} object\footnote{An alternative would be to use the recently proposed \code{BigInt} support for JavaScript (\url{https://github.com/tc39/proposal-bigint}), but this feature is currently only available in Chrome.}, enabling an analysis to faithfully reason about 64-bit integers.

\section{Implementation}
\label{sec:implementation}

We have implemented the Wasabi instrumenter, including the static analyses it performs, in about 5000 lines of Rust code.
Rust programs themselves can be compiled to WebAssembly, which gives us the option to run Wasabi in the browser and instrument WebAssembly programs at load time in the future.
To reduce the time required for instrumenting large binaries, Wasabi can instrument multiple WebAssembly functions in parallel.
The only synchronization point is the map of low-level hooks created during on-demand monomorphization, which is guarded by an upgradeable multiple readers/single writer lock.

Our implementation is available to the public under the permissive MIT license at
\url{http://wasabi.software-lab.org}.

\section{Evaluation}
\label{sec:evaluation}

To evaluate Wasabi, we focus on five research questions:
\begin{enumerate}[label=\itshape\bfseries RQ\,\arabic*,labelsep=1em,leftmargin=*,topsep=\smallskipamount,itemsep=\smallskipamount]
	\item How easy is it to write dynamic analyses with Wasabi?
	\item Do the instrumented WebAssembly programs remain faithful to the original execution?
	\item How long does it take to instrument programs?
	\item How much does the code size increase?
	\item What is the runtime overhead due to instrumentation?
\end{enumerate}

\subsection{Experimental Setup}

We apply Wasabi to \nbBenchmarkPrograms{} programs.
30 of them are from the PolyBench/C benchmark suite\footnote{\url{http://web.cse.ohio-state.edu/~pouchet.2/software/polybench/}}, which has been used in the paper introducing WebAssembly~\cite{Haas:2017:BWU:3062341.3062363}.
In total, the PolyBench benchmark suite comprises 5,163 non-empty, non-comment lines of C code.
We compile the PolyBench programs to WebAssembly with emscripten 1.38.8, resulting in 790\,KB of WebAssembly binaries.
Moreover, we use two complex, real-world programs:
the Unreal Engine~4 Zen Garden demo\footnote{\url{https://s3.amazonaws.com/mozilla-games/ZenGarden/EpicZenGarden.html}}, as an example of a major game engine running in the browser, and the PSPDFKit benchmark\footnote{\url{https://pspdfkit.com/webassembly-benchmark/}}, which exercises a commercial library for in-browser rendering and annotation of PDFs.
Their WebAssembly binaries are 39.5\,MB and 9.5\,MB, respectively.

All experiments are performed on a laptop with an Intel Core i7-7500U CPU (2 cores, hyper-threading, 2.7 to 3.5\,GHz, 4\,MB L3 cache) and 16\,GB of RAM. The operating system is Ubuntu 17.10 64-bit. To execute WebAssembly programs, we use a nightly version of Firefox 63.0a1 (2018-08-02).%

\subsection{Ease of Implementing Analyses (RQ\,1)}
\label{sec:eval-analyses}

\begin{table}
	\setlength{\tabcolsep}{2pt}
	\begin{adjustbox}{center}
		\begin{tabular}{@{}l@{\hspace*{-2pt}}rr@{}}
			\toprule
			Analysis & Hooks & LOC\\
			\midrule
			Instruction mix analysis & all & 42\\
			Basic block profiling & \code{begin} & 9\\[\medskipamount]
			Instruction coverage & all & 11\\
			Branch coverage & \code{if}, \code{br\_if}, \code{br\_table}, \code{select} & 14\\[\medskipamount]
			Call graph analysis & \code{call\_pre} & 18\\
			Dynamic taint analysis & all & 208\\
			Cryptominer detection & \code{binary} & 10\\
			Memory access tracing & \code{load}, \code{store} & 11\\
			\bottomrule
		\end{tabular}
	\end{adjustbox}
	\caption{Analyses built on top of Wasabi.}
	\label{tab:analyses}
\end{table}

We have implemented \nbAnalyses{} dynamic analyses on top of Wasabi.
Table~\ref{tab:analyses} lists them, along with the hooks they implement and their total lines of JavaScript code.

\paragraph{Instruction Mix Analysis}
This analysis counts how often each kind of instruction is executed, which can serve as a basis for performance and security analyses.

\paragraph{Basic Block Profiling}
A classic dynamic analysis~\cite{chang1988trace} that counts how often each function, block, and loop is executed, which is useful, e.g., for finding \enquote{hot} code.

\paragraph{Instruction and Branch Coverage}
These analyses record for each instruction and branch, respectively, whether it is executed, which is useful to assess the quality of tests.

\paragraph{Call Graph Analysis}
This analysis creates a dynamic call graph, including indirect calls and calls between functions that are neither imported nor exported.
Call graphs are the basis of various other analyses, e.g., to find dynamically dead code or to reverse-engineer malware.

\paragraph{Taint Analysis}
The analysis associates a taint with every value and tracks how taints propagate through instructions, function calls, and memory accesses, to detect illegal flows from sources to sinks.
	
\paragraph{Memory Access Tracing}
The analysis tracks all memory accesses and stores them for a later off-line analysis, e.g., to detect cache-unfriendly access patterns.
	
\paragraph{Cryptominer Detection}
As discussed in the introduction, this analysis gathers a signature based on the frequency of binary instructions to identify mining of cryptocurrencies~\cite{SEISMIC}.
	
\begin{figure}
\begin{lstlisting}[language=JavaScript]
const coverage = [];
function addBranch({func, instr}, branch) {
    coverage[func] = coverage[func] || [];
    coverage[func][instr] = coverage[func][instr] || [];
    if (!coverage[func][instr].includes(branch)) {
        coverage[func][instr].push(branch);
    }
}
Wasabi.analysis = {
    if_(loc, cond) { addBranch(loc, cond) },
    br_if(loc, target, cond) { addBranch(loc, cond) },
    br_table(loc, tbl, df, idx) { addBranch(loc, idx) },
    select(loc, cond) { addBranch(loc, cond) },
};
\end{lstlisting}
	\caption{Branch coverage analysis with Wasabi.}
	\label{fig:branch-coverage}
\end{figure}

\medskip\noindent
As illustrated by the low numbers of lines of code in Table~\ref{tab:analyses}, each of these analyses can be implemented with little effort.
For further illustration, \Cref{fig:branch-coverage} shows the implementation of the branch coverage analysis.
It implements four hooks, \code{if}, \code{br\_if}, \code{br\_table}, and \code{select} to keep track of all branches.

\subsection{Faithfulness of Execution (RQ\,2)}
\label{sec:eval-faithfulness}

To validate that Wasabi's instrumentation does not modify the semantics of the original program, we compare the behavior of each unmodified binary with the behavior of the fully instrumented binary.
For the PolyBench programs, we compile each program with an option to output intermediate results of every calculation on the console.
Similarly, the Unreal Engine demo has a mode to check that the pixel values of rendered frames are the same as pre-defined reference frames.
For all these programs, the behavior remains unchanged after instrumentation.
The PSPDFKit benchmark does not provide any built-in correctness check; based on our manual observations the behavior of the original and instrumented code appear to be the same.

As another way to validate the instrumented WebAssembly code, we use the static WebAssembly validator, which offers some well-formedness guarantees and checks that the code is type correct~\cite{WasmSpec}.
Running \code{wasm-validate} from the WebAssembly Binary Toolkit\footnote{\url{https://github.com/WebAssembly/wabt}} on all 32 fully instrumented programs shows that all the instrumented code passes the validator.
We also instrument and successfully validate Wasabi's output on all programs of the official WebAssembly specification test suite\footnote{\url{https://github.com/WebAssembly/spec/tree/master/test}}, which consists of 63 additional programs.

\subsection{Time to Instrument (RQ\,3)}
\label{sec:eval-time-instrument}

\begin{table}
	\begin{adjustbox}{center}
	\begin{tabular}{l@{\hspace*{3pt}}rr@{\ }r@{\ }rr}
		\toprule
		Program & Binary Size (B) & \multicolumn{3}{r}{Runtime (ms)} & $\frac{\text{MB}}{\text{s}}$\\
		\midrule
		PolyBench (avg.) & 26\,332 $\pm$ 299 & 23 & $\pm$ & 1.4 & 1.15 \\
		PSPDFKit & 9\,615\,389 & 5\,129 & $\pm$ & 65 & 1.87 \\
		Unreal Engine 4 & 39\,510\,398 & 15\,481 & $\pm$ & 293 & 2.55\\
		\bottomrule
	\end{tabular}
	\end{adjustbox}
	\caption{Time taken to instrument programs, averaged across 20 runs (and 30 programs for PolyBench).}
	\label{tab:runtime}
\end{table}

\Cref{tab:runtime} shows how long Wasabi takes to instrument the programs. The $x \pm y$ notation means a mean value of $x$ and a standard deviation of $y$ after 20 repetitions.
For readability, we have summarized the results for all 30 PolyBench programs in one row.
While the PolyBench programs are of similar, small size (26.3\,KB\,$\pm$\,299\,B), the PSPDFKit and Unreal Engine binaries are considerably larger (9.6\,MB and 39.5\,MB, respectively).
Instrumentation takes 23ms, on average, for the PolyBench programs, i.e., it is almost instantaneous, and still quick for the larger PSPDFKit (5s) and Unreal Engine binaries (15.5s).
Wasabi's instrumentation is parallelized (\Cref{sec:implementation}), and these numbers are obtained with four threads running on two physical cores.
The single-threaded instrumentation time on the large Unreal Engine binary is on average 26.5s, showing that the parallelization reduces the execution time to 15.5/26.5 $\approx$ 0.58 of the single-threaded time.
The last column of \Cref{tab:runtime} reports the throughput, i.e., binary code processed per second, showing that the throughput increases with larger binary sizes.

\subsection{Increase of Code Size (RQ\,4)}
\label{sec:eval-code-size}

\begin{figure*}
	\includegraphics[width=\textwidth]{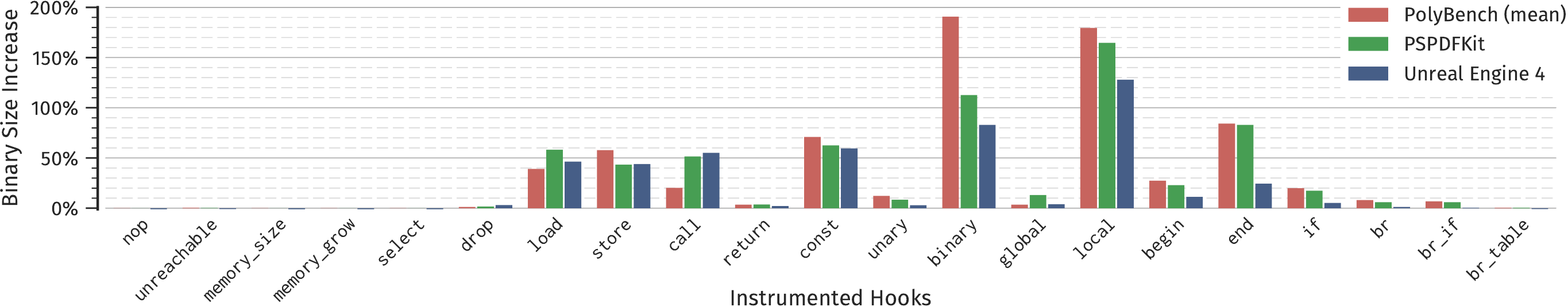}
	\caption{Binary size increase in percent of the original size, when instrumenting the test programs for different analysis hooks. For readability, binary sizes for the 30 PolyBench programs are shown averaged.}
	\label{fig:binary-size}
	
	\includegraphics[width=\textwidth,trim=0 0 0 -10pt]{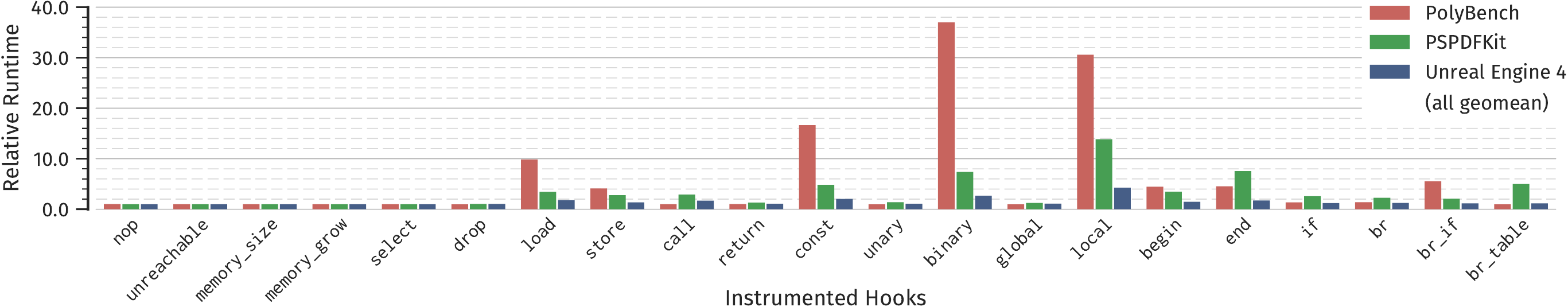}
	\caption{Runtime of the instrumented programs relative to the uninstrumented runtime, per analysis hook. Results are averaged over 20 runs (and again for readability, over the 30 PolyBench programs).}
	\label{fig:runtime-analysis}
\end{figure*}

\Cref{fig:binary-size} presents the increase in binary code size after instrumenting a program.
Since many analyses need only a small subset of all hooks (e.g., block profiling needs only \texttt{begin}), we evaluate code size increase per required hook, as provided by selective instrumentation (\Cref{sec:selectiveInstr}).
For each hook on the x-axis, the figure shows on the y-axis the increase in binary size as a percentage of the original program size.
That is, 0\% means the instrumented binary has the same size as the original one and 100\% means the program doubled in size due to instrumentation.

With selective instrumentation, more than half of the hooks increase the binary size only by a negligible amount or not at all (less than 1\% increase for \texttt{nop}, \texttt{unreachable}, \texttt{memory\_size}, \texttt{memory\_grow}, \texttt{select}, and \texttt{br\_table}; less than 10\% for \texttt{drop}, \texttt{return}, \texttt{unary}, \texttt{global}, \texttt{if}, \texttt{br}, and \texttt{br\_if}, on average). In fact, in several cases the Unreal Engine binary decreased by 1\% because Wasabi encodes indices more compactly than the original binary.\footnote{WebAssembly uses the variable length LEB128 encoding for integers, also known from the DWARF debug information format~\cite{DWARFSpec}. This allows for multiple possible encodings with different lengths of the same number.}

Naturally, hooks for instructions that appear very often in the program have the largest influence on the code size, e.g., memory \texttt{load} and \texttt{store} (between 39\% and 58\% increase), \texttt{begin} and \texttt{end} of blocks (11\%\,--\,84\%), pushing to the stack with \texttt{const} (59\,--\,71\%), operations on \texttt{local}s (128\,--\,180\%), and finally \texttt{binary} instructions (83\,--\,190\%).
The difference for the \texttt{binary} hook between PolyBench and the other programs can be explained by the former being mostly numerical computation (thus having more binary instructions such as \code{i32.mul}), whereas PSPDFKit and the Unreal Engine have more diverse code.
When instrumenting for all hooks together, which is not required for many analyses, the size increases between 495 and 743\%. This result shows that selective instrumentation is very effective in reducing the binary size, compared to blindly instrumenting all instructions.

To evaluate Wasabi's on-demand monomorphization of hooks, we count how many low-level hooks are inserted during full instrumentation.
For PolyBench, between 110 (\code{floyd-warshall} program) and 122 (\code{deriche}) hooks are inserted, 302 hooks for PSPDFKit, and 783 hooks for the Unreal Engine.
In the original Unreal Engine binary, i.e., a real-world WebAssembly program, the call with the largest number of arguments passes 22 \code{i32} values, which clearly shows that eagerly generating all possible monomorphic combinations of call hooks ($4^{22} \approx 1.7 \times 10^{13}$) is simply not possible. 
Even in the small PolyBench programs, calls to functions with 6 arguments are common.
For these programs, generating no more than 122 hooks on-demand is much better than generating all $4^6 = 4,096$ hooks for \code{call} instructions plus some more for other instructions.

\subsection{Runtime Overhead (RQ\,5)}
\label{sec:eval-runtime-overhead}

\Cref{fig:runtime-analysis} shows how much runtime overhead the instrumentation imposes.
On the y-axis, we show the runtime of the instrumented program relative to the original program, that is, a value of 1.0x means the runtime does not increase due to instrumentation.
As for code size, most of the hooks contribute only a small runtime overhead: \code{nop}, \code{unreachable}, \code{memory\_size}, \code{memory\_grow}, \code{select}, \code{drop}, and \code{unary} each impose less than 1.02x overhead, on average.
Instrumenting for \code{return} or \code{call} hooks, which are sufficient for many interesting analyses at the function level, incurs a reasonable overhead of up to 1.3x or 2.8x, respectively.
More expensive hooks are \texttt{begin} and \texttt{end} for observing blocks, which incur between 1.5x and 9.9x runtime overhead, 1.8x\,--\,20x for \texttt{load}, up to 6.5x for \texttt{store}, 2x\,--\,32x for \texttt{const}, 4x\,--\,48.5x for operations on \texttt{local}s, and 2.6x\,--\,77.5x for \texttt{binary} operations.
When instrumenting for all hooks, the runtime overhead is between 49x and 163x. 
Note that the overheads for the PolyBench programs, which perform only numerical computations, are much higher than for the real-world workloads in PSPDFKit and the Unreal Engine.
Typical WebAssembly programs call out to the host environment, e.g., to perform shading in WebGL, modify the DOM, or interact with some other Web API, so any overhead imposed by Wasabi contributes only to parts of the total execution time.

Comparing the overhead results to existing heavy-weight dynamic analysis frameworks for other languages shows that Wasabi's overhead is reasonable.
The widely-used JavaScript analysis framework Jalangi reports overheads with the \emph{empty} analysis in the same order of magnitude, namely 26x during record plus 30x during replay, on average \cite{Sen:2013:JSR:2491411.2491447}.
Similarly, the RoadRunner analysis framework for JVM byte code reports an average slowdown of 52x without any analysis \cite{Flanagan:2010:RDA:1806672.1806674}.

\section{Related Work}

\paragraph{WebAssembly}

WebAssembly has been first publicly announced in 2015 and since 2017 is implemented by four major browser engines (Chrome, Edge, Firefox, and Safari).
A paper by Haas et al.~\cite{Haas:2017:BWU:3062341.3062363} formalizes the language and its type system, and explains the design rationales.
Watt describes the mechanized, formal verification of the WebAssembly specification \cite{Watt:2018:MVW:3176245.3167082}.
Herrera et al.\ study the performance of WebAssembly, compared to JavaScript, for numerical benchmarks~\cite{herrera2018webassembly}.

\paragraph{Dynamic Analysis of WebAssembly}
Despite its young age, several dynamic analyses for WebAssembly have already been proposed, including two taint analyses~\cite{2018arXiv180201050F,szanto2018taint} and a cryptomining detector~\cite{SEISMIC}.
These analyses have been implemented by modifying the V8 engine~\cite{2018arXiv180201050F},
by implementing a new WebAssembly virtual machine in JavaScript~\cite{szanto2018taint}, and
through custom binary instrumentation~\cite{SEISMIC}, respectively.
Our evaluation shows that these analyses and others can be implemented in top of Wasabi with significantly less effort.

\paragraph{Binary Instrumentation Tools}
Binary instrumentation has been a popular strategy to implement dynamic analyses.
Often used tools for x86 binaries include
DynamoRIO~\cite{Bruening:2003:IAD:776261.776290},
Pin~\cite{Luk:2005:PBC:1065010.1065034}, and
Valgrind~\cite{Nethercote:2007:VFH:1250734.1250746}, which have provided inspiration for Wasabi.
These tools instrument binaries at runtime by translating basic blocks just before their execution, and by storing translations in a code cache.
In contrast, Wasabi instruments binaries statically, i.e., ahead-of-time, which avoids any instrumentation overhead during execution.
Wasabi also differs w.r.t.\ the API it provides to analysis authors: While DynamoRIO provides an API to manipulate instructions, Wasabi provides an API to observe the execution of instructions.
Analyses written for Pin can specify ``instrumentation routines'', which determine where to place calls to analysis routines. 
Instead, Wasabi automatically selects which kinds of instructions to instrument based on the hooks implemented by the analysis.
Umbra~\cite{DBLP:conf/cgo/ZhaoBA10} is a dynamic binary instrumentation tool that focuses on efficient memory shadowing.
In contrast, Wasabi provides a general-purpose framework for arbitrary dynamic analysis, including memory shadowing.
A difference compared to all the above tools is that in Wasabi, the dynamic analysis is written and executed in a high-level language, JavaScript, instead of being compiled to binary code.
The rationale is that JavaScript is already very popular in the web, making it easier for analysis authors to adopt Wasabi.

\paragraph{Dynamic Analysis in General}

Dynamic analysis~\cite{ball1999concept} has since long been recognized as an effective way to complement static analysis~\cite{ernst2003static}.
Various analyses have been proposed, including dynamic slicing~\cite{Agrawal1990},
taint analyses for x86 binaries~\cite{Newsome2005} and Android~\cite{Enck:2014:TIT:2642648.2619091},
tools to find concurrency bugs~\cite{Lu2006,Park2009,Burckhardt2010a} and 
heap-related bugs~\cite{Chilimbi2006},
an analysis to track the origin of null and undefined values~\cite{Bond2007},
and analyses to understand performance problems~\cite{Yu2014}.
Given the increasing interest in WebAssembly, we expect an increased demand for dynamic analyses for WebAssembly, for which Wasabi provides a reusable platform.

\paragraph{Dynamic Analysis for the Web}
Motivated by the dynamic features of JavaScript, such as runtime loading of code, various dynamic analyses for JavaScript-web applications have been presented.
For example, recent approaches include analyses to find type inconsistencies~\cite{icse2015}, JIT-unfriendly code~\cite{fse2015}, bad coding practices~\cite{issta2015}, and data races~\cite{Petrov2012}.
Many of these analysis are built on top of Jalangi~\cite{Sen:2013:JSR:2491411.2491447}, a general-purpose dynamic analysis framework for JavaScript.
To the best of our knowledge, there is no comparable tool for WebAssembly yet, a gap this paper aims to fill.

\section{Conclusion}

This paper presents Wasabi, a general-purpose dynamic analysis framework for WebAssembly, the new low-level instruction set for the web.
The framework instruments binaries ahead-of-time and inserts code into the binary that calls into an analysis implemented in JavaScript.
Besides being the first dynamic analysis framework for WebAssembly, Wasabi addresses several unique challenges that did not occur in dynamic analysis tools for other platforms.
In particular, we handle the problem of tracing polymorphic instructions with analysis functions that have a fixed type via an on-demand monomorphization of analysis hooks,
and we statically resolve relative branch labels in control-flow constructs during the instrumentation.
The high-level API provided to analyses authors allows for implementing otherwise complex analyses with a few dozens of lines of code, while still providing a complete view of the execution.
Our evaluation with both compute-intensive benchmark programs and real-world web applications shows 1.02x to 163x runtime overhead, depending on the program and which instructions are analyzed, which is reasonable for heavyweight dynamic analyses.

We believe that Wasabi provides a solid basis for various analyses to be implemented in the future.
As an interesting challenge for future work, we envision cross-language dynamic analysis, in particular, to analyze web applications that run both JavaScript and WebAssembly code.

\bibliographystyle{ACM-Reference-Format}
{%
\bibliography{references}
}

\end{document}